\documentclass[journal]{IEEEtran}

\usepackage{amsmath,amssymb,amsfonts,amsthm}

\usepackage{graphicx}
\usepackage{array}
\usepackage{makecell}
\usepackage{booktabs}
\usepackage[caption=false,font=normalsize,labelfont=sf,textfont=sf]{subfig}

\usepackage{algorithm2e}

\usepackage{textcomp}
\usepackage{stfloats}
\usepackage{url}
\usepackage{verbatim}
\usepackage{cite}
\usepackage{xcolor}

\newcounter{assumption}

\newcounter{theorm}

\begin{document}
	
	\title{Pinching-Antenna-Assisted Integrated AirComp–NOMA: Joint Transceiver and Antenna-Position Optimization
    }


	\author{IEEE Publication Technology,~\IEEEmembership{Staff,~IEEE,}
	}

	\author{Saeid Pakravan, Wessam Ajib, Wei-Ping Zhu 

\thanks{S. Pakravan and W. Ajib are with the Department of Computer Sciences, University of Quebec in Montreal (UQAM), Montreal, QC, Canada. email: pakravan.saeid@uqam.ca; ajib.wessam@uqam.ca.}


\thanks{W.-P. Zhu is with the Department of Electrical and Computer Engineering, Concordia University, Montreal, QC, Canada. email: weiping@ece.concordia.ca.}



}
	\maketitle

\begin{abstract}
Integrating over-the-air computation (AirComp) with non-orthogonal
multiple access (NOMA) enables concurrent data aggregation and
information transmission, but tightly couples computation accuracy
with communication quality of service (QoS). This paper investigates
a pinching-antenna-assisted uplink, where flexible pinching-antenna
(PA) placement provides additional spatial degrees of freedom to
manage this coupling. We minimize the AirComp mean-squared error (MSE) subject to NOMA QoS, transmit-power, and PA deployment constraints by jointly designing the AirComp transceiver, PA positions, and successive interference cancellation (SIC) order.
Exploiting the recursive structure of uplink SIC, we establish an
exact NOMA feasibility condition and derive an optimal SIC-ordering
rule. These structural results eliminate the NOMA power variables and
the combinatorial search over decoding orders, reducing the original
problem to a continuous optimization problem. We then develop an iterative algorithm with optimal AirComp transmit-coefficient and receive-scaling updates and a feasibility-preserving majorization scheme for PA placement. The proposed algorithm preserves NOMA feasibility while generating a monotonically nonincreasing computation-MSE sequence. Numerical results demonstrate the computation-accuracy gains of jointly optimizing the AirComp transceiver and PA positions under NOMA QoS requirements.
\end{abstract}


\section{Introduction}
\label{sec:introduction}

The convergence of wireless communication and distributed intelligence is reshaping the design objectives of next-generation networks. Emerging applications such as federated learning, distributed sensing, and edge intelligence require wireless devices not only to exchange individual information but also to efficiently aggregate distributed data \cite{ref_edge_intelligence,
ref_aircomp_survey, 11342412}. Supporting these heterogeneous communication and computation services through separate transmission mechanisms can incur considerable communication overhead and latency, motivating more integrated use of shared wireless resources.

Over-the-air computation (AirComp) provides an efficient mechanism for wireless data aggregation by exploiting the waveform superposition property of multiple-access channels \cite{ref_aircomp_fundamental, 10092857}. Rather than individually recovering the data transmitted by participating devices, AirComp directly reconstructs a desired function from their superimposed signals, thereby reducing the communication overhead associated with conventional transmit-then-compute protocols. This capability has motivated extensive research on transceiver design, power control, and channel reconfiguration for improving computation accuracy \cite{ref_aircomp_transceiver,ref_aircomp_ris, 11517417,
ref_aircomp_ma}. In practical wireless networks, however, computation-oriented devices may coexist with users requiring individual information delivery. Integrating AirComp with non-orthogonal multiple access (NOMA) enables these heterogeneous services to coexist over the same time-frequency resources \cite{ref_integrated_aircomp_noma}. Such integration introduces an intrinsic communication–computation coupling: the signals used for AirComp affect the decoding conditions of NOMA users, whereas their quality-of-service (QoS) requirements constrain the resources available for accurate computation.

Additional spatial degrees of freedom offer a promising means of
managing this coupling. Pinching antenna systems (PASs) employ pinching antennas (PAs) whose positions can be configured along dielectric waveguides
\cite{ref_pas_architecture,ref_pas_overview, 11626907}. Unlike conventional fixed-position antenna arrays, adjusting the PA positions changes the free-space propagation distances as well as the attenuation and phase accumulation along the waveguide, thereby enabling spatial reconfiguration of the effective wireless channels \cite{11625963}. This capability has stimulated recent
research on PAS-assisted NOMA, including joint PA-position, power,
beamforming, and resource-allocation designs
\cite{ref_pas_noma_qos,ref_pas_noma_jointbf,ref_pas_noma_ra}.
PASs have also been investigated for AirComp, where the additional
spatial flexibility can be exploited to improve computation accuracy
\cite{ref_pas_aircomp,ref_pas_aircomp_programmable}.

Despite these advances, the joint design of AirComp, NOMA, and configurable PA placement has received limited attention. Integrated AirComp–NOMA designs enable simultaneous data aggregation and individual information transmission but do not exploit PA-enabled channel reconfiguration \cite{ref_integrated_aircomp_noma}. Existing PAS-assisted AirComp studies optimize computation performance without accounting for the QoS and successive interference cancellation (SIC) requirements of coexisting NOMA users \cite{ref_pas_aircomp,ref_pas_aircomp_programmable}. Conversely, recent PAS–NOMA studies focus on communication-oriented objectives through PA-enabled spatial design and communication-resource optimization \cite{ref_pas_noma_qos,ref_pas_noma_jointbf,ref_pas_noma_ra}. Related movable-antenna research has also considered joint antenna-position, power-control, and SIC-order optimization in uplink NOMA \cite{ref_ma_noma}. More recently in \cite{ref_conference}, we investigated a PAS-assisted hybrid AirComp–NOMA uplink with joint precoding and PA placement from a computation–communication rate perspective. However, the SIC order was determined based on channel gains, without exploiting the underlying NOMA structure.
A summary of the most relevant related works and their key distinctions is provided in Table~\ref{tab:lit_comparison}. Nevertheless, the computation-centric joint design of the AirComp transceiver, PA positions, and SIC order under per-user NOMA QoS requirements introduces a distinct set of coupled design challenges. In particular, PA placement simultaneously affects the AirComp distortion and NOMA feasibility, while the latter further depends on the SIC decoding order. This coupling gives rise to a mixed continuous-discrete optimization problem in which the AirComp transceiver and PA positions are intertwined with NOMA feasibility and the combinatorial SIC decoding order.

\begin{table*}[t]
\centering
\caption{Comparison of the Proposed Work With Representative Related Studies}
\label{tab:lit_comparison}
\renewcommand{\arraystretch}{1.1}
\setlength{\tabcolsep}{3.1pt}
\footnotesize
\begin{tabular}{|p{0.7cm}|p{1.9cm}|c|c|c|c|c|p{4.9cm}|p{3cm}|}
\hline
\textbf{Ref.} &
\textbf{Performance Metric} &
\multicolumn{5}{c|}{\textbf{System and Design Considerations}} &
\textbf{Main Contribution} &
\textbf{Solution Approach} \\
\cline{3-7}
&
&
\textbf{AirComp} &
\textbf{NOMA} &
\textbf{PAS} &
\textbf{Comm. QoS} &
\textbf{SIC Order} &
&
\\
\hline

\cite{ref_integrated_aircomp_noma} &
Computation MSE &
\checkmark &
\checkmark &
-- &
\checkmark &
-- &
Integrated AirComp--NOMA design for simultaneous data aggregation
and information transmission under a communication-rate requirement &
Analytical and iterative optimization
\\
\hline

\cite{ref_pas_noma_qos} &
Communication rate &
-- &
\checkmark &
\checkmark &
\checkmark &
-- &
Joint PA-position and power-allocation design under heterogeneous
QoS requirements &
BCD / SCA
\\
\hline

\cite{ref_pas_noma_jointbf} &
Transmit power &
-- &
\checkmark &
\checkmark &
\checkmark &
-- &
Joint transmit and pinching-beamforming design for NOMA &
MM / PDD-based optimization
\\
\hline

\cite{ref_pas_noma_ra} &
Sum rate &
-- &
\checkmark &
\checkmark &
\checkmark &
-- &
Joint power allocation, pinching beamforming, and user scheduling &
PDD / BCD / matching
\\
\hline

\cite{ref_pas_aircomp} &
Computation MSE &
\checkmark &
-- &
\checkmark &
-- &
-- &
Joint PA placement, transmit-power, and receive-processing design
for AirComp &
Alternating optimization
\\
\hline

\cite{ref_pas_aircomp_programmable} &
Computation MSE &
\checkmark &
-- &
\checkmark &
-- &
-- &
PAS-assisted AirComp design with programmable aggregation &
Joint optimization
\\
\hline
\cite{ref_ma_noma} &
Sum rate &
-- &
\checkmark &
-- &
-- &
\checkmark &
Joint movable-antenna-position, power-control, and SIC-order design &
SCA / closed-form optimization
\\
\hline

\cite{ref_conference} &
Computation--communication rate &
\checkmark &
\checkmark &
\checkmark &
\checkmark &
-- &
Joint precoding and PA-placement design for hybrid AirComp--NOMA &
Alternating optimization
\\
\hline

\textbf{This Work} &
\textbf{Computation MSE} &
\checkmark &
\checkmark &
\checkmark &
\checkmark &
\checkmark &
\textbf{Joint AirComp transceiver, PA-position, and SIC-order design
with exact NOMA feasibility characterization} &
\textbf{Analytical reduction and feasibility-preserving
majorization}
\\
\hline
\end{tabular}
\\ \vspace{1ex}
\scriptsize
{
\textbf{BCD}: block coordinate descent; \textbf{SCA}:
successive convex approximation; \textbf{MM}: majorization--minimization;
\textbf{PDD}: penalty dual decomposition.
}
\end{table*}

The main contributions of this paper are summarized as follows.
\begin{itemize}

\item
We formulate a computation-centric optimization problem for a PAS-assisted integrated AirComp–NOMA uplink, jointly optimizing the AirComp transceiver, NOMA transmit coefficients, PA positions, and SIC decoding order to minimize the AirComp MSE under transmit-power, NOMA QoS, and PA-deployment constraints.

\item
We establish an exact necessary-and-sufficient NOMA feasibility condition for arbitrary SIC orders, characterize the componentwise minimum NOMA transmit powers, and derive an optimal SIC-ordering rule. These results eliminate the NOMA power variables and combinatorial decoding-order search, reducing the original problem to a continuous optimization problem.

\item
We develop an iterative optimization algorithm combining optimal block-wise AirComp transceiver updates with feasibility-preserving majorization for PA placement. The algorithm maintains NOMA feasibility and guarantees a monotonically nonincreasing, convergent MSE sequence.

\item
Numerical results demonstrate the AirComp-MSE improvements over benchmark schemes and quantify the effects of NOMA QoS requirements, PA deployment, and waveguide attenuation on computation accuracy.

\end{itemize}

The remainder of this paper is organized as follows.
Section~\ref{sec:system_model} presents the system model and problem
formulation. Section~\ref{sec:proposed_framework} develops the
proposed optimization framework and analyzes its convergence and
computational complexity. Section~IV provides the numerical results,
and Section~V concludes the paper.

{\bf{Notation:}} Throughout this paper, scalars, vectors, and sets are denoted by
italic letters, boldface lowercase letters, and calligraphic letters,
respectively. The operators
$(\cdot)^*$, $|\cdot|$, $\|\cdot\|_2$, $\arg(\cdot)$, $\operatorname{Re}\left\{ . \right\}$,
and $\mathbb{E}[\cdot]$ denote complex conjugation, absolute value,
Euclidean norm, phase, real part, and expectation, respectively.
$\mathcal{CN}(\mu,\sigma^2)$ denotes a circularly symmetric complex
Gaussian distribution with mean $\mu$ and variance $\sigma^2$, and
$\nabla f(\mathbf{v})$ denotes the gradient of a real-valued function
$f$ with respect to the real vector $\mathbf{v}$. Finally, $\triangleq$ denotes equality by definition, and
$\Pi(\mathcal{S})$ denotes the set of all permutations of a finite set
$\mathcal{S}$.

\section{System Model and Problem Formulation}
\label{sec:system_model}

\subsection{Network Architecture}
\label{subsec:network_architecture}

\begin{figure}[]
    \centering
\includegraphics[width=1\linewidth, height=0.5\linewidth]{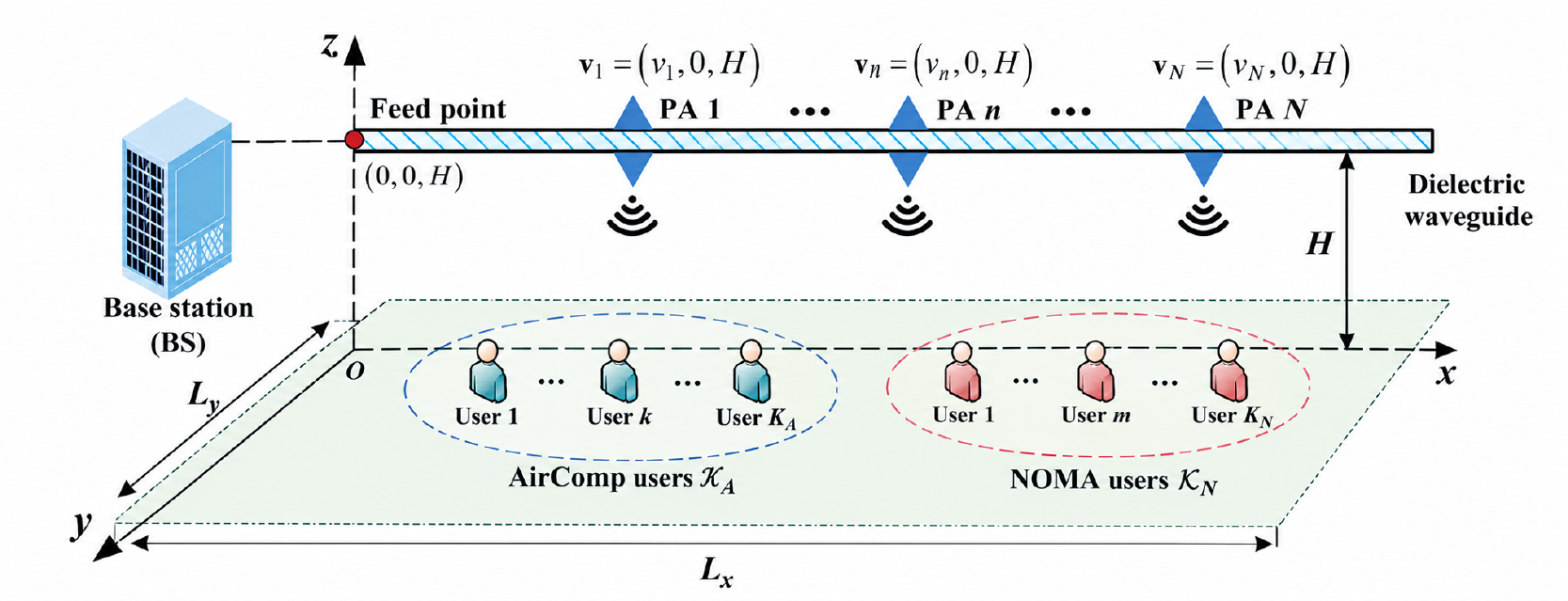}
    \caption{Illustration of the considered PAS-assisted integrated AirComp–NOMA uplink system.}
    \label{fig:system_model}
\end{figure}

As illustrated in Fig.~\ref{fig:system_model}, we consider the uplink of a
PAS-assisted integrated AirComp--NOMA system, where a base station (BS)
equipped with a PAS simultaneously serves computation-oriented and
communication-oriented users over the same time-frequency resources.
The computation-oriented users employ AirComp
for wireless aggregation of function-related data, whereas the
communication-oriented users employ power-domain NOMA to transmit
individual information-bearing signals to the BS.

Let $\mathcal K_{\rm A}\triangleq\{1,\ldots,K_{\rm A}\}$ and
$\mathcal K_{\rm N}\triangleq\{1,\ldots,K_{\rm N}\}$ denote the sets of
AirComp and NOMA users, respectively, with
$\mathcal K_{\rm A}\cap\mathcal K_{\rm N}=\emptyset$, and define
$\mathcal U\triangleq\mathcal K_{\rm A}\cup\mathcal K_{\rm N}$ as the
set of all users.

A three-dimensional Cartesian coordinate system is adopted to characterize
the network geometry. The users are located within a rectangular
$L_x\times L_y$ service region on the horizontal plane, with user
$u\in\mathcal U$ located at $\mathbf q_u=(x_u,y_u,0)$. The user
locations are assumed to be known at the BS during the considered
transmission interval.

The BS employs a PAS comprising $N$ pinching antennas (PAs) deployed
along a dielectric waveguide of length $L_x$, installed parallel to the
$x$-axis at height $H$. The waveguide feed point is located at
$\mathbf v_0=(0,0,H)$, while the $n$-th PA is located at
$\mathbf v_n=(v_n,0,H)$, where $v_n$ denotes its adjustable position
along the waveguide. The PAs are indexed according to their positions,
such that $0\leq v_1<\cdots<v_N\leq L_x$, and adjacent PAs satisfy the
minimum-spacing requirement $v_n-v_{n-1}\geq D_{\min}$,
$n=2,\ldots,N$, where $D_{\min}>0$ denotes the minimum allowable
inter-PA distance. The PA positions are collected in
$\mathbf v\triangleq[v_1,\ldots,v_N]^T$.

\subsection{PAS-Aided Uplink Channel Model}
\label{subsec:pas_channel_model}

We consider a line-of-sight (LoS)-dominant propagation model between each
user and the PAs \cite{ref_conference, 11481959, 11595855}. The distance between user $u\in\mathcal U$ and the
$n$-th PA is given by
\begin{equation}
d_{u,n}(v_n)
\triangleq
\|\mathbf q_u-\mathbf v_n\|
=
\sqrt{(x_u-v_n)^2+y_u^2+H^2}.
\label{eq:user_pa_distance}
\end{equation}
Accordingly, the free-space channel coefficient between user $u$ and the
$n$-th PA is modeled as
\begin{equation}
h_{u,n}(v_n)
=
\frac{\lambda}{4\pi d_{u,n}(v_n)}
\exp\!\left(
-j\frac{2\pi}{\lambda}d_{u,n}(v_n)
\right),
\label{eq:physical_channel}
\end{equation}
where $\lambda$ denotes the free-space carrier wavelength.

After being coupled into the dielectric waveguide, the signal received
through each PA propagates toward the feed point and experiences both
attenuation and phase accumulation. Accordingly, the guided-wave response
associated with the $n$-th PA is modeled as
\begin{equation}
\phi_n(v_n)
=
\exp(-\alpha_{\rm g}v_n)
\exp\!\left(
-j\frac{2\pi}{\lambda_{\rm g}}v_n
\right),
\label{eq:guided_wave_response}
\end{equation}
where $\alpha_{\rm g}\geq0$ denotes the in-waveguide attenuation
coefficient and
$
    \lambda_{\rm g}
    =
    \frac{\lambda}{n_{\rm eff}}
$
denotes the guided wavelength, with $n_{\rm eff}$ representing the
effective refractive index of the dielectric waveguide.

The equivalent uplink channel between user $u$ and the PAS feed point is
therefore obtained by coherently combining the contributions received
through all PAs as
\begin{equation}
g_u(\mathbf v)
=
\sum_{n=1}^{N}
h_{u,n}(v_n)\phi_n(v_n).
\label{eq:equivalent_channel}
\end{equation}
Equivalently, substituting
\eqref{eq:physical_channel} and
\eqref{eq:guided_wave_response} into
\eqref{eq:equivalent_channel} yields
\begin{equation}
\begin{aligned}
g_u(\mathbf v)
=
\sum_{n=1}^{N}
&
\frac{\lambda}
{4\pi d_{u,n}(v_n)}
\exp(-\alpha_{\rm g}v_n)
\\
&\times
\exp\!\left[
-j\left(
\frac{2\pi}{\lambda}d_{u,n}(v_n)
+
\frac{2\pi}{\lambda_{\rm g}}v_n
\right)
\right].
\end{aligned}
\label{eq:equivalent_channel_expanded}
\end{equation}
For notational simplicity, the dependence of $g_u(\mathbf{v})$ on the
PA-position vector $\mathbf{v}$ is omitted whenever no ambiguity arises.

\subsection{Uplink Signal Model}
\label{subsec:signal_model}

Let $s_k\in\mathbb C$, $k\in\mathcal K_{\rm A}$, denote the
preprocessed data symbol transmitted by AirComp user $k$, and let
$d_j\in\mathbb C$, $j\in\mathcal K_{\rm N}$, denote the
information-bearing symbol transmitted by NOMA user $j$. All transmitted
symbols are assumed to be mutually independent, zero-mean, and normalized
to unit power, i.e.,
$\mathbb E[|s_k|^2]=\mathbb E[|d_j|^2]=1$.

AirComp user $k$ applies a complex transmit coefficient $p_k$, whereas
NOMA user $j$ employs a transmit coefficient $p_j$. The corresponding
individual power constraints are
\begin{equation}
    |p_k|^2\le P_k,\quad k\in\mathcal K_{\rm A},
    \qquad
    |p_j|^2\le P_j,\quad j\in\mathcal K_{\rm N},
\label{eq:power_constraints}
\end{equation}
where $P_k$ and $P_j$ denote the respective maximum transmit powers.

Using the equivalent PAS channel in \eqref{eq:equivalent_channel}, the
aggregate baseband signal observed at the PAS feed-point receiver is
given by
\begin{equation}
y =
\sum_{k\in\mathcal K_{\rm A}} g_k(\mathbf v)p_k s_k
+
\sum_{j\in\mathcal K_{\rm N}} g_j(\mathbf v)p_j d_j
+ z,
\label{eq:received_signal}
\end{equation}
where $z\sim\mathcal{CN}(0,\sigma^2)$ denotes the effective additive
Gaussian noise at the receiver, with $\sigma^2$ being the corresponding
noise power.

\subsection{NOMA Reception and Achievable Rates}
\label{subsec:noma_reception}

The BS first decodes the NOMA signals via successive interference
cancellation (SIC), while treating the simultaneously transmitted AirComp
signals as interference. Let
$\boldsymbol{\pi}=[\pi(1),\ldots,\pi(K_{\rm N})]$ denote the SIC
decoding order of the NOMA users, where user $\pi(1)$ is decoded first
and user $\pi(K_{\rm N})$ is decoded last.

Under ideal SIC, when decoding user $\pi(m)$, the signals of users
$\pi(1),\ldots,\pi(m-1)$ have already been decoded and canceled,
whereas those of users $\pi(m+1),\ldots,\pi(K_{\rm N})$ remain as
intra-NOMA interference. Since the AirComp signals carry function-related
symbols intended for over-the-air aggregation rather than individual
decoding, they are not canceled during the NOMA decoding stage. Their
aggregate interference power is therefore
\begin{equation}
I_{\rm A}
\triangleq
\sum_{k\in\mathcal K_{\rm A}}
|g_k|^2 |p_k|^2 .
\label{eq:aircomp_interference}
\end{equation}
Accordingly, the signal-to-interference-plus-noise ratio (SINR) for decoding user $\pi(m)$ is
\begin{equation}
\Gamma_{\pi(m)}
=
\frac{
|g_{\pi(m)}|^2 |p_{\pi(m)}|^2
}{
\displaystyle
\sum_{\ell=m+1}^{K_{\rm N}}
|g_{\pi(\ell)}|^2 |p_{\pi(\ell)}|^2
+
I_{\rm A}
+
\sigma^2
},
\quad m=1,\ldots,K_{\rm N}.
\label{eq:noma_sinr}
\end{equation}
Assuming Gaussian signaling, the corresponding achievable rate is
\begin{equation}
R_{\pi(m)}
=
B\log_2\!\left(1+\Gamma_{\pi(m)}\right),
\quad m=1,\ldots,K_{\rm N},
\label{eq:noma_rate}
\end{equation}
where $B$ denotes the system bandwidth.

\subsection{AirComp Reception and Computation MSE}
\label{subsec:aircomp_reception}

After successfully decoding and canceling all NOMA signals, the residual
signal at the BS is
\begin{equation}
y_{\rm A}
=
\sum_{k\in\mathcal K_{\rm A}} g_k p_k s_k + z.
\label{eq:aircomp_received_signal}
\end{equation}
The BS aims to recover the arithmetic sum of the AirComp users' data
symbols, i.e.,
$
s
\triangleq
\sum_{k\in\mathcal K_{\rm A}} s_k .
\label{eq:desired_sum}
$
To this end, a complex receive scaling coefficient $w$ is applied to
$y_{\rm A}$, yielding
$
\hat{s}
=
w^* y_{\rm A}.
\label{eq:aircomp_estimate}
$
The computation accuracy is characterized by the MSE
between the estimated and desired sums. Using the independence and
unit-power assumptions of the transmitted symbols, the computation MSE
is given by
\begin{equation}
\mathrm{MSE}
\triangleq \mathbb{E}\!\left[|\hat{s}-s|^2\right] 
= \sum_{k\in\mathcal{K}_{\rm A}}
\left|w^*g_kp_k-1\right|^2
+\sigma^2|w|^2.
\label{eq:AirComp_MSE}
\end{equation}

\subsection{Problem Formulation}
\label{subsec:problem_formulation}

We aim to minimize the AirComp MSE while guaranteeing the
QoS requirements of all NOMA users. Accordingly, the transmit
coefficients, the receive scaling coefficient, the PA positions, and
the SIC decoding order are jointly designed as
\begin{subequations}
\label{prob:P0}
\begin{align}
\mathcal P_0:
\underset{\{p_u\}_{u\in\mathcal U},\,w,\,\mathbf v,\,
\boldsymbol{\pi}}
{\operatorname{minimize}}
\quad &
\sum_{k\in\mathcal K_{\rm A}}
\left|w^*g_k(\mathbf v)p_k-1\right|^2
+\sigma^2|w|^2
\label{prob:P0_obj}
\\
\operatorname{subject~to}\quad
&
R_{\pi(m)} \ge R_{\pi(m)}^{\min},
\quad m=1,\ldots,K_{\rm N},
\label{prob:P0_rate}
\\
&
|p_u|^2 \le P_u,
\quad u\in\mathcal U,
\label{prob:P0_power}
\\
&
\boldsymbol{\pi}\in\Pi(\mathcal K_{\rm N}),
\label{prob:P0_order}
\\
&
0 \leq v_1 < \cdots < v_N \leq L_x, 
\label{prob:P0_region}
\\
&
v_n-v_{n-1}\ge D_{\min},
\quad n=2,\ldots,N.
\label{prob:P0_spacing}
\end{align}
\end{subequations}
where $\mathbf v=[v_1,\ldots,v_N]^T$, $P_u$ denotes the maximum
transmit power of user $u$, and $\Pi(\mathcal K_{\rm N})$ denotes the
set of all permutations of the NOMA-user set $\mathcal K_{\rm N}$.
Constraint~\eqref{prob:P0_rate} guarantees the prescribed minimum
communication QoS of each NOMA user,
\eqref{prob:P0_power} imposes the individual transmit-power budgets,
and \eqref{prob:P0_order} specifies a valid SIC decoding order.
Constraints~\eqref{prob:P0_region} and \eqref{prob:P0_spacing} confine
the PAs to the physical extent of the dielectric waveguide and enforce
the minimum separation between adjacent PAs, respectively.

Problem~$\mathcal P_0$ is challenging due to the discrete SIC decoding
order, the coupling among the transmit coefficients, the receive
scaling coefficient, and the PA positions, and the nonlinear
dependence of the AirComp MSE and NOMA QoS constraints on the
position-dependent equivalent channels. The next section exploits the
structure of the NOMA QoS constraints and SIC decoding order to reduce
the dimensionality and combinatorial complexity of $\mathcal P_0$
before addressing the remaining continuous optimization variables.

\section{Proposed Optimization Framework}
\label{sec:proposed_framework}

We exploit the structure of the NOMA QoS constraints to eliminate the transmit powers for a given SIC decoding order and then characterize the order that maximizes the admissible interference-plus-noise level. This eliminates both the NOMA power variables and the combinatorial search over decoding orders, reducing the original joint design to an optimization over the remaining continuous variables.

\subsection{NOMA Feasibility Characterization and Power Elimination}
\label{subsec:noma_feasibility}

For a given SIC decoding order $\boldsymbol{\pi}$, let
$\gamma_j\triangleq 2^{R_j^{\min}/B}-1>0$ denote the minimum required
SINR of NOMA user $j\in\mathcal K_{\rm N}$, where
$R_j^{\min}>0$. Moreover, define the aggregate interference-plus-noise
term as
$T\triangleq I_{\rm A}+\sigma^2
=\sum_{k\in\mathcal K_{\rm A}}
|g_k(\mathbf v)|^2|p_k|^2+\sigma^2$.

\textit{Proposition 1:}
For given $\mathbf v$, $\{p_k\}_{k\in\mathcal K_{\rm A}}$, and
$\boldsymbol{\pi}$, there exists a NOMA power allocation satisfying
both the QoS and individual transmit-power constraints if and only if
\begin{equation}
T
\leq
\frac{
|g_{\pi(m)}(\mathbf v)|^2 P_{\pi(m)}
}{
\gamma_{\pi(m)}
\displaystyle\prod_{\ell=m+1}^{K_{\rm N}}
\left(1+\gamma_{\pi(\ell)}\right)
},
\quad
m=1,\ldots,K_{\rm N}.
\label{eq:noma_feasibility_condition}
\end{equation}
Whenever \eqref{eq:noma_feasibility_condition} holds, the
componentwise minimum NOMA transmit-power allocation is given by
\begin{equation}
q_{\pi(m)}^{\min}
=
\frac{
\gamma_{\pi(m)}T
}{
|g_{\pi(m)}(\mathbf v)|^2
}
\prod_{\ell=m+1}^{K_{\rm N}}
\left(1+\gamma_{\pi(\ell)}\right),
\quad
m=1,\ldots,K_{\rm N},
\label{eq:min_noma_power}
\end{equation}
where $q_j\triangleq |p_j|^2$, and the empty product for
$m=K_{\rm N}$ is defined as one.

\textit{Proof:}
See Appendix~\ref{app:proof_prop1}.

Proposition~1 allows the NOMA transmit powers to be eliminated from
the joint optimization. For any SIC decoding order
$\boldsymbol{\pi}\in\Pi(\mathcal K_{\rm N})$, define
\begin{equation}
\Theta_{\boldsymbol{\pi}}(\mathbf v)
\triangleq
\min_{m=1,\ldots,K_{\rm N}}
\frac{
|g_{\pi(m)}(\mathbf v)|^2 P_{\pi(m)}
}{
\gamma_{\pi(m)}
\displaystyle\prod_{\ell=m+1}^{K_{\rm N}}
\left(1+\gamma_{\pi(\ell)}\right)
}.
\label{eq:noma_tolerance}
\end{equation}
The quantity $\Theta_{\boldsymbol{\pi}}(\mathbf v)$ represents the
maximum aggregate interference-plus-noise level that can be tolerated
by all NOMA users under the decoding order $\boldsymbol{\pi}$ while
satisfying their QoS requirements and individual power budgets.
Therefore, by Proposition~1, the NOMA QoS and power constraints in
$\mathcal P_0$ are jointly equivalent to
$
I_{\rm A}+\sigma^2
\leq
\Theta_{\boldsymbol{\pi}}(\mathbf v).
$
Using \eqref{eq:aircomp_interference}, Problem~$\mathcal P_0$ can thus
be equivalently reduced to
\begin{subequations}
\label{prob:P1}
\begin{align}
\mathcal P_1:
\underset{\{p_k\}_{k\in\mathcal K_{\rm A}},\,w,\,\mathbf v,\,
\boldsymbol{\pi}}
{\operatorname{minimize}}
\quad &
\sum_{k\in\mathcal K_{\rm A}}
\left|w^*g_k(\mathbf v)p_k-1\right|^2
+\sigma^2|w|^2
\label{prob:P1_obj}
\\
\operatorname{subject~to}\
&
\sum_{k\in\mathcal K_{\rm A}}
|g_k(\mathbf v)|^2|p_k|^2+\sigma^2
\leq
\Theta_{\boldsymbol{\pi}}(\mathbf v),
\label{prob:P1_noma}
\\
&
|p_k|^2\leq P_k,
\quad k\in\mathcal K_{\rm A},
\label{prob:P1_power}
\\
&
\boldsymbol{\pi}\in\Pi(\mathcal K_{\rm N}),
\label{prob:P1_order}
\\
&
0 \leq v_1,\quad v_N \leq L_x,
\label{prob:P1_region}
\\
&
v_n-v_{n-1}\geq D_{\min},
\quad n=2,\ldots,N.
\label{prob:P1_spacing}
\end{align}
\end{subequations}

\subsection{Optimal SIC Decoding Order and Problem Reduction}
\label{subsec:sic_order}

Although the NOMA transmit powers have been eliminated in
$\mathcal P_1$, the SIC decoding order remains a discrete optimization
variable. We next show that, for any given PA-position vector
$\mathbf v$, an optimal decoding order can be obtained directly
without enumerating all $K_{\rm N}!$ permutations. Define the ordering
metric
\[
\rho_j(\mathbf v)
\triangleq
\frac{1+\gamma_j}{\gamma_j}
|g_j(\mathbf v)|^2 P_j,
\qquad j\in\mathcal K_{\rm N}.
\]

\textit{Proposition 2:}
For any given $\mathbf v$, the maximum NOMA interference tolerance
over all SIC decoding orders is attained by any permutation
$\boldsymbol{\pi}^{\star}(\mathbf v)\in\Pi(\mathcal K_{\rm N})$
satisfying
\begin{equation}
\rho_{\pi^\star(1)}(\mathbf v)
\geq
\rho_{\pi^\star(2)}(\mathbf v)
\geq
\cdots
\geq
\rho_{\pi^\star(K_{\rm N})}(\mathbf v).
\label{eq:optimal_sic_order}
\end{equation}
Equivalently,
\begin{equation}
\Theta_{\boldsymbol{\pi}^{\star}(\mathbf v)}(\mathbf v)
=
\max_{\boldsymbol{\pi}\in\Pi(\mathcal K_{\rm N})}
\Theta_{\boldsymbol{\pi}}(\mathbf v).
\label{eq:max_noma_tolerance}
\end{equation}

\textit{Proof:}
See Appendix~\ref{app:proof_prop2}.

Proposition~2 eliminates the need for an exhaustive search over the
SIC decoding orders. For any fixed $\mathbf v$, the objective of
$\mathcal P_1$ is independent of $\boldsymbol{\pi}$, while the
decoding order affects feasibility only through
$\Theta_{\boldsymbol{\pi}}(\mathbf v)$ in
\eqref{prob:P1_noma}. Therefore, selecting an order that maximizes $\Theta_{\boldsymbol{\pi}}(\mathbf v)$
yields the largest admissible AirComp interference level and is
optimal for the given $\mathbf{v}$.

Based on Proposition~2, define the optimal NOMA interference tolerance
for a given PA-position vector $\mathbf v$ as
\begin{equation}
\Theta^\star(\mathbf v)
\triangleq
\max_{\boldsymbol{\pi}\in\Pi(\mathcal K_{\rm N})}
\Theta_{\boldsymbol{\pi}}(\mathbf v)
=
\Theta_{\boldsymbol{\pi}^\star(\mathbf v)}(\mathbf v),
\label{eq:optimal_noma_tolerance}
\end{equation}
where $\boldsymbol{\pi}^\star(\mathbf v)$ is obtained according to
\eqref{eq:optimal_sic_order}. Substituting
\eqref{eq:optimal_noma_tolerance} into $\mathcal P_1$ eliminates the
SIC decoding-order variable and yields the equivalent reduced problem
\begin{subequations}
\label{prob:P2}
\begin{align}
\mathcal P_2:\quad
\underset{\{p_k\}_{k\in\mathcal K_{\rm A}},\,w,\,\mathbf v}
{\operatorname{minimize}}
\quad &
\sum_{k\in\mathcal K_{\rm A}}
\left|w^*g_k(\mathbf v)p_k-1\right|^2
+\sigma^2|w|^2
\label{prob:P2_obj}
\\
\operatorname{subject~to}\quad
&
\sum_{k\in\mathcal K_{\rm A}}
|g_k(\mathbf v)|^2|p_k|^2+\sigma^2
\leq
\Theta^\star(\mathbf v),
\label{prob:P2_noma}
\\
&
|p_k|^2\leq P_k,
\quad k\in\mathcal K_{\rm A},
\label{prob:P2_power}
\\
&
0 \leq v_1,\quad v_N \leq L_x,
\label{prob:P2_region}
\\
&
v_n-v_{n-1}\geq D_{\min},
\quad n=2,\ldots,N.
\label{prob:P2_spacing}
\end{align}
\end{subequations}

Problem~$\mathcal P_2$ is equivalent to $\mathcal P_0$ in terms of
the minimum achievable AirComp MSE, while the NOMA
transmit powers and the SIC decoding order have been eliminated as
explicit optimization variables. After solving $\mathcal P_2$, an
optimal SIC decoding order is obtained from
\eqref{eq:optimal_sic_order} at the optimized PA positions, and the
corresponding minimum NOMA transmit powers are recovered from
\eqref{eq:min_noma_power}.

\subsection{Optimal Receive Scaling Coefficient}
\label{subsec:receive_scaling}

For fixed AirComp transmit coefficients and PA positions, the
constraints of $\mathcal P_2$ are independent of $w$. Hence, the
receive scaling coefficient can be optimized independently. Expanding
the AirComp MSE with respect to $w$ yields
\begin{equation}
\begin{aligned}
\operatorname{MSE}(w)
={}&
\left(
\sum_{k\in\mathcal K_{\rm A}}
|g_k(\mathbf v)p_k|^2+\sigma^2
\right)|w|^2
\\
&-2\operatorname{Re}\!\left\{
w^*
\sum_{k\in\mathcal K_{\rm A}}
g_k(\mathbf v)p_k
\right\}
+K_{\rm A}.
\end{aligned}
\label{eq:mse_w}
\end{equation}
Since $\sigma^2>0$, \eqref{eq:mse_w} is strictly convex in $w$.
Setting its Wirtinger derivative with respect to $w^*$ to zero gives
\begin{equation}
\left(
\sum_{k\in\mathcal K_{\rm A}}
|g_k(\mathbf v)p_k|^2+\sigma^2
\right)w
=
\sum_{k\in\mathcal K_{\rm A}}
g_k(\mathbf v)p_k .
\label{eq:w_stationarity}
\end{equation}
Therefore, the unique optimal receive scaling coefficient is
\begin{equation}
w^\star
=
\frac{
\displaystyle\sum_{k\in\mathcal K_{\rm A}}
g_k(\mathbf v)p_k
}{
\displaystyle\sum_{k\in\mathcal K_{\rm A}}
|g_k(\mathbf v)p_k|^2+\sigma^2
}.
\label{eq:optimal_receive_scaling}
\end{equation}

\subsection{Optimal AirComp Transmit Coefficients}
\label{subsec:aircomp_coefficients}

We next optimize the AirComp transmit coefficients for fixed $w$ and
$\mathbf v$. Define the available AirComp interference budget as
\begin{equation}
Q(\mathbf v)
\triangleq
\Theta^\star(\mathbf v)-\sigma^2.
\label{eq:aircomp_budget}
\end{equation}
For any PA-position vector admitting a feasible solution of
$\mathcal P_2$, $Q(\mathbf v)\geq0$. Since $\sigma^2|w|^2$ is
constant in this subproblem, it can be omitted from the objective.

For $w\neq0$ and $g_k(\mathbf v)\neq0$, the objective is minimized,
for any given $|p_k|$, by phase-aligning $w^*g_k(\mathbf v)p_k$ with
the desired AirComp coefficient. Hence,
$\arg(p_k^\star)=\arg(w)-\arg(g_k(\mathbf v))$.
Define the received amplitude
$r_k\triangleq|g_k(\mathbf v)p_k|$ and
$\bar r_k\triangleq|g_k(\mathbf v)|\sqrt{P_k}$. The
transmit-coefficient optimization then reduces to
\begin{subequations}
\label{prob:aircomp_amplitude}
\begin{align}
\underset{\{r_k\}}
{\operatorname{minimize}}
\quad &
\sum_{k\in\mathcal K_{\rm A}}
\left(|w|r_k-1\right)^2
\label{prob:aircomp_amplitude_obj}
\\
\operatorname{subject~to}\quad
&
\sum_{k\in\mathcal K_{\rm A}}r_k^2
\leq Q(\mathbf v),
\label{prob:aircomp_amplitude_budget}
\\
&
0\leq r_k\leq\bar r_k,
\quad k\in\mathcal K_{\rm A}.
\label{prob:aircomp_amplitude_bound}
\end{align}
\end{subequations}

For $w\neq0$, Problem~\eqref{prob:aircomp_amplitude} is convex with a
strictly convex objective and therefore admits a unique optimal
received-amplitude vector. If $Q(\mathbf v)=0$, the solution is
$r_k^\star=0$ for all $k\in\mathcal K_{\rm A}$. For
$Q(\mathbf v)>0$, the Karush–Kuhn–Tucker (KKT) conditions yield
\begin{equation}
r_k^\star
=
\min\left\{
\bar r_k,\,
\frac{|w|}{|w|^2+\mu^\star}
\right\},
\quad k\in\mathcal K_{\rm A},
\label{eq:optimal_received_amplitude}
\end{equation}
where $\mu^\star\geq0$ is the multiplier associated with
\eqref{prob:aircomp_amplitude_budget}. In particular,
$\mu^\star=0$ if
$\sum_{k\in\mathcal K_{\rm A}}
\min\{\bar r_k,1/|w|\}^2\leq Q(\mathbf v)$.
Otherwise, the interference-budget constraint is active, and the
unique $\mu^\star>0$ satisfies
\begin{equation}
\sum_{k\in\mathcal K_{\rm A}}
\left[
\min\left\{
\bar r_k,\,
\frac{|w|}{|w|^2+\mu^\star}
\right\}
\right]^2
=
Q(\mathbf v),
\label{eq:aircomp_dual_equation}
\end{equation}
which can be efficiently obtained by bisection.

Finally, for $w\neq0$ and $g_k(\mathbf v)\neq0$, the optimal AirComp
transmit coefficient is recovered as
\begin{equation}
p_k^\star
=
\frac{r_k^\star}{|g_k(\mathbf v)|}
e^{j[\arg(w)-\arg(g_k(\mathbf v))]} .
\label{eq:optimal_aircomp_coefficient}
\end{equation}
If $w=0$ or $g_k(\mathbf v)=0$, $p_k^\star=0$ can be selected without
loss of optimality.

\subsection{PA-position Optimization}
\label{subsec:pa_optimization}

We finally optimize the PA positions for fixed $w$ and
$\{p_k\}_{k\in\mathcal K_{\rm A}}$. Directly handling
$\Theta^\star(\mathbf v)$ in \eqref{prob:P2_noma} is inconvenient
because the optimal SIC order may change with $\mathbf v$. We instead
exploit Propositions~1 and~2 to construct a feasibility-preserving
PA-position update.

At the current iterate $\mathbf v^{(t)}$, let
$\boldsymbol{\pi}^{(t)}
=\boldsymbol{\pi}^{\star}(\mathbf v^{(t)})$
be obtained from \eqref{eq:optimal_sic_order}, and define
\begin{equation}
a_m^{(t)}
\triangleq
\frac{P_{\pi^{(t)}(m)}}
{\gamma_{\pi^{(t)}(m)}
\displaystyle\prod_{\ell=m+1}^{K_{\rm N}}
\left(1+\gamma_{\pi^{(t)}(\ell)}\right)},
\quad m=1,\ldots,K_{\rm N}.
\label{eq:pa_noma_coefficient}
\end{equation}
For the fixed order $\boldsymbol{\pi}^{(t)}$, Proposition~1 shows
that the NOMA QoS and power constraints are feasible if and only if
\begin{equation}
c_m^{(t)}(\mathbf v)
\triangleq
\sum_{k\in\mathcal K_{\rm A}}
|p_k|^2|g_k(\mathbf v)|^2+\sigma^2
-a_m^{(t)}
|g_{\pi^{(t)}(m)}(\mathbf v)|^2
\leq0.
\label{eq:pa_noma_constraint}
\end{equation}
The PA-dependent AirComp objective is
\begin{equation}
f(\mathbf v)
\triangleq
\sum_{k\in\mathcal K_{\rm A}}
\left|w^*p_k g_k(\mathbf v)-1\right|^2 .
\label{eq:pa_objective}
\end{equation}

Although \eqref{eq:pa_noma_constraint} and
\eqref{eq:pa_objective} are generally nonconvex in $\mathbf v$, they
are continuously differentiable over the feasible PA region. We
therefore construct the quadratic upper models
\begin{equation}
\widehat f_t(\mathbf v)
={}
f(\mathbf v^{(t)})
+\nabla f(\mathbf v^{(t)})^T
(\mathbf v-\mathbf v^{(t)})
+
\frac{L_f^{(t)}}{2}
\|\mathbf v-\mathbf v^{(t)}\|_2^2,
\label{eq:pa_objective_majorizer}
\end{equation}
and
\begin{equation}
\widehat c_{m,t}(\mathbf v)
={}
c_m^{(t)}(\mathbf v^{(t)})
+\nabla c_m^{(t)}(\mathbf v^{(t)})^T
(\mathbf v-\mathbf v^{(t)})
+
\frac{L_m^{(t)}}{2}
\|\mathbf v-\mathbf v^{(t)}\|_2^2,
\label{eq:pa_constraint_majorizer}
\end{equation}
where the positive curvature parameters $L_f^{(t)}$ and
$L_m^{(t)}$ are selected by backtracking to satisfy the required
upper-bound conditions. The corresponding gradients and
majorization procedure are detailed in
Appendix~\ref{app:pa_majorization}.

The PA positions are then updated by solving the convex quadratically constrained quadratic program (QCQP)
\begin{subequations}
\label{prob:pa_sca}
\begin{align}
\mathbf v^{(t+1)}
\in\arg\min_{\mathbf v}\quad&
\widehat f_t(\mathbf v)
\label{prob:pa_sca_obj}
\\
\operatorname{subject~to}\quad&
\widehat c_{m,t}(\mathbf v)\leq0,
\quad m=1,\ldots,K_{\rm N},
\label{prob:pa_sca_noma}
\\
&
0\leq v_1,\quad v_N\leq L_x,
\label{prob:pa_sca_region}
\\
&
v_n-v_{n-1}\geq D_{\min},
\quad n=2,\ldots,N.
\label{prob:pa_sca_spacing}
\end{align}
\end{subequations}
The surrogate functions are tight at $\mathbf v^{(t)}$, while their
upper-bound properties ensure that the accepted update remains
feasible for $\boldsymbol{\pi}^{(t)}$ and does not increase the
AirComp objective. After obtaining $\mathbf v^{(t+1)}$, the SIC order
is updated according to Proposition~2. Since
$\Theta_{\boldsymbol{\pi}^{\star}(\mathbf v^{(t+1)})}
(\mathbf v^{(t+1)})
\geq
\Theta_{\boldsymbol{\pi}^{(t)}}(\mathbf v^{(t+1)})$,
this reordering preserves the NOMA feasibility established by the
PA-position update.

The resulting joint optimization procedure is summarized in Algorithm~\ref{alg:proposed_algorithm}.

\RestyleAlgo{ruled}
\SetNlSty{textbf}{}{:}
\begin{algorithm}[t]
\SetAlgoLined

\textbf{Initialize:} Select a feasible PA-position vector
$\mathbf v^{(0)}$ satisfying $Q(\mathbf v^{(0)})>0$, choose
$w^{(0)}\neq0$, and set $t=0$\;

\Repeat{the convergence criterion is satisfied}{
    Determine the optimal SIC order
    $\boldsymbol{\pi}^{(t)}
    =\boldsymbol{\pi}^{\star}(\mathbf v^{(t)})$
    according to \eqref{eq:optimal_sic_order}\;

    Update the AirComp transmit coefficients
    $\{p_k^{(t+1)}\}_{k\in\mathcal K_{\rm A}}$
    using \eqref{eq:optimal_received_amplitude}--%
    \eqref{eq:optimal_aircomp_coefficient}\;

    Update the receive scaling coefficient $w^{(t+1)}$
    using \eqref{eq:optimal_receive_scaling}\;

    Update the PA positions $\mathbf v^{(t+1)}$
    by solving \eqref{prob:pa_sca} for
    $\boldsymbol{\pi}^{(t)}$ with backtracking\;

    Set $t\leftarrow t+1$\;
}

Recover the final SIC order and NOMA transmit powers using
\eqref{eq:optimal_sic_order} and \eqref{eq:min_noma_power},
respectively\;

\caption{Proposed Joint Optimization Algorithm}
\label{alg:proposed_algorithm}
\end{algorithm}

\subsection{Convergence and Complexity Analysis}
\label{subsec:convergence_complexity}

The AirComp transmit-coefficient and receive-scaling updates globally
solve their respective subproblems, while the accepted PA-position
update does not increase the objective, as established in
Appendix~\ref{app:pa_majorization}. Hence, the MSE is monotonically
nonincreasing. Feasibility is also preserved: the AirComp update
satisfies the interference-budget constraint, the receive-scaling
update does not affect NOMA feasibility, and the PA-position update
preserves feasibility for the current SIC order. Moreover, by
Proposition~2, updating the SIC order cannot decrease the admissible
interference-plus-noise level. Since the MSE is nonnegative, the MSE
sequence is bounded below and therefore converges.

The computational complexity of each outer iteration is analyzed as
follows. Computing the equivalent channels for all users requires
$\mathcal O(N(K_{\rm A}+K_{\rm N}))$ operations, while determining
the optimal SIC order by sorting requires
$\mathcal O(K_{\rm N}\log K_{\rm N})$. The AirComp
transmit-coefficient update requires
$\mathcal O((I_{\mu}+1)K_{\rm A})$ operations, where $I_{\mu}$
denotes the number of bisection iterations, whereas the closed-form
receive-scaling update requires $\mathcal O(K_{\rm A})$. The
PA-position update solves the convex QCQP \eqref{prob:pa_sca} with
$N$ variables, $K_{\rm N}$ quadratic constraints, and linear
PA-position constraints. Denoting by
$\mathcal C_{\rm QCQP}(N,K_{\rm N})$ the cost of one QCQP solve and
by $I_{\rm bt}$ the number of solves required by backtracking, the
overall complexity per outer iteration is
$\mathcal O\big(N(K_{\rm A}+K_{\rm N})
+K_{\rm N}\log K_{\rm N}
+(I_{\mu}+1)K_{\rm A}
+I_{\rm bt}\mathcal C_{\rm QCQP}(N,K_{\rm N})\big)$,
which is typically dominated by the PA-position update.

\section{Numerical Results}
\label{sec:numerical_results}

This section evaluates the proposed joint design and provides
insights into the interplay among AirComp accuracy, NOMA
QoS, SIC ordering, and PA deployment.

\subsection{Simulation Setup}
\label{subsec:simulation_setup}

Following representative settings in related studies
\cite{ref_pas_aircomp,ref_pas_noma_qos,11481959, 11595855, yang2026noma,11643214}, we set
the carrier frequency and system bandwidth to $f_c=28$ GHz and
$B=1$ MHz, respectively. The dielectric waveguide has length
$L_x=10$ m and is deployed at height $H=3$ m, while the users are
independently and uniformly distributed over a $10\times10$ m$^2$
service region. The effective refractive index is
$n_{\rm eff}=1.4$, and the minimum PA spacing is
$D_{\min}=\lambda/2$. The in-waveguide attenuation is $0.08$ dB/m,
corresponding to $\alpha_g=9.21\times10^{-3}$ m$^{-1}$ in the
channel model.
We consider $K_{\rm A}=4$ AirComp users, $K_{\rm N}=3$ NOMA users,
and $N=6$ PAs. Unless otherwise stated, all users have a maximum
transmit power of $P_{\max}=20$ dBm, each NOMA user requires
$R^{\min}=0.5$ Mbit/s, and the effective receiver noise power is
$\sigma^2=-90$ dBm.

The algorithm terminates when the relative MSE change between two
successive outer iterations is below $10^{-4}$ or after $100$
iterations. Unless otherwise stated, the results are averaged over
$500$ independent user-location realizations. The default simulation
parameters are summarized in Table~\ref{tab:simulation_parameters}.

\begin{table}[t]
\centering
\caption{Default Simulation Parameters}
\label{tab:simulation_parameters}
\renewcommand{\arraystretch}{1.12}
\setlength{\tabcolsep}{5pt}
\footnotesize
\begin{tabular}{l c}
\hline
\textbf{Parameter} & \textbf{Value} \\
\hline
Carrier frequency, $f_c$                   & $28$ GHz \\
System bandwidth, $B$                      & $1$ MHz \\
Service region, $L_x\times L_y$            & $10\times10$ m$^2$ \\
Waveguide length, $L_x$                    & $10$ m \\
PAS height, $H$                            & $3$ m \\
Effective refractive index, $n_{\rm eff}$  & $1.4$ \\
Attenuation coefficient, $\alpha_g$        & $9.21\times10^{-3}$ m$^{-1}$ \\
Minimum PA spacing, $D_{\min}$             & $\lambda/2$ \\
Number of PAs, $N$                         & $6$ \\
Number of AirComp users, $K_{\rm A}$       & $4$ \\
Number of NOMA users, $K_{\rm N}$          & $3$ \\
Maximum transmit power, $P_{\max}$         & $20$ dBm \\
NOMA rate requirement, $R^{\min}$          & $0.5$ Mbit/s \\
Effective receiver noise power, $\sigma^2$ & $-90$ dBm \\
Number of realizations                     & $500$ \\
\hline
\end{tabular}
\end{table}

\subsection{Benchmark Schemes}

For comparison, we consider the following benchmark schemes:
\begin{itemize}
    \item {\bf{Fixed-PA:}} The PAs are uniformly deployed along the waveguide as
    $v_n=nL_x/(N+1)$, while the remaining design variables are optimized as in the proposed scheme.

    \item {\bf{Random-PA:}} The PA positions are randomly generated for each user realization and kept fixed, while the remaining variables are optimized as in the proposed scheme.

    \item {\bf{Discrete-PA:}} The PA positions are restricted to a predefined set of feasible locations along the waveguide, while the remaining design variables are optimized as in the proposed scheme.

    \item {\bf{Equal-Power AirComp:}} All AirComp users employ a common transmit-power level, selected as the largest value satisfying the individual power constraints and NOMA feasibility condition, while the PA positions, SIC order, transmit phases, and receive scaling factor are optimized as in the proposed scheme.
\end{itemize}

\begin{figure}[]
    \centering
\includegraphics[width=0.9\linewidth, height=0.6\linewidth]{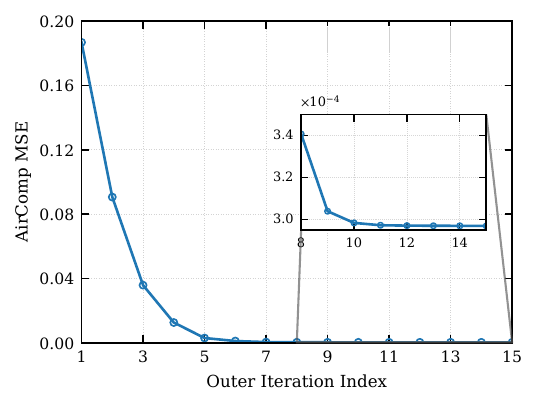}
    \caption{Convergence behavior of the proposed algorithm.}
    \label{fig:convergence}
\end{figure}

Fig.~\ref{fig:convergence} illustrates the convergence behavior of the proposed algorithm. The AirComp MSE decreases monotonically over the iterations, exhibiting a rapid reduction in the initial iterations and gradually approaching a stable value. This behavior corroborates the convergence analysis and
demonstrates rapid numerical convergence.

\begin{figure}[]
    \centering
\includegraphics[width=0.9\linewidth, height=0.55\linewidth]{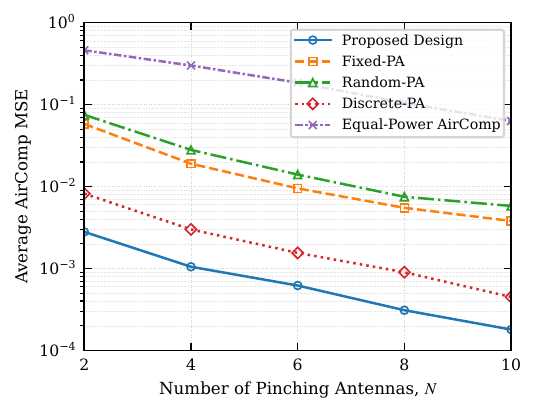}
    \caption{Average AirComp MSE versus the number of pinching antennas.}
    \label{fig:mse_vs_N}
\end{figure}

Fig.~\ref{fig:mse_vs_N} illustrates the AirComp MSE versus the number of PAs. 
The proposed design consistently achieves the lowest MSE, with further improvement as $N$ increases, demonstrating the benefit of jointly optimizing the PA positions and AirComp transceiver. 
The gaps over the Fixed-PA and Random-PA schemes confirm the importance of adaptive PA positioning, while the performance of Discrete-PA reveals the loss incurred by restricting the PAs to predefined locations. 
Moreover, the substantially higher MSE of Equal-Power AirComp highlights the importance of user-specific power optimization in the integrated AirComp--NOMA system.

\begin{figure}[]
    \centering
\includegraphics[width=0.9\linewidth, height=0.6\linewidth]{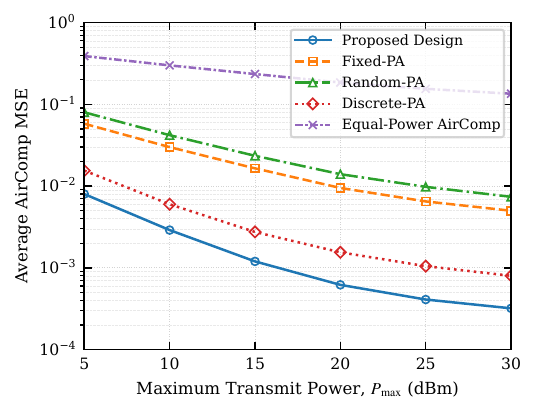}
    \caption{Average AirComp MSE versus the maximum transmit power.}
    \label{fig:mse_vs_power}
\end{figure}

Fig.~\ref{fig:mse_vs_power} illustrates the impact of the maximum transmit power on the AirComp MSE. 
The MSE decreases with $P_{\max}$ for all schemes, as a larger power budget provides greater flexibility for signal alignment while satisfying the NOMA QoS constraints. 
The proposed design consistently achieves the lowest MSE, with clear gains over the fixed, random, and discrete PA deployments. 
Moreover, the pronounced gap from Equal-Power AirComp highlights the benefit of user-specific power optimization.

\begin{figure}[]
    \centering
\includegraphics[width=0.9\linewidth, height=0.6\linewidth]{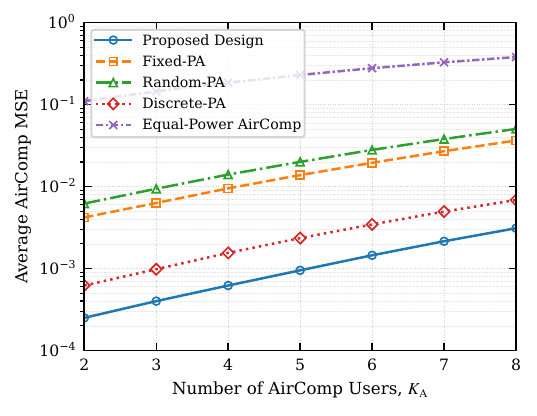}
    \caption{Average AirComp MSE versus the number of AirComp users.}
    \label{fig:mse_vs_KA}
\end{figure}

Fig.~\ref{fig:mse_vs_KA} illustrates the impact of the number of
AirComp users on the computation MSE. As $K_{\mathrm A}$ increases,
the MSE generally increases because a larger number of distributed
signals must be simultaneously aligned for function aggregation,
while their aggregate interference also tightens the NOMA QoS
constraints. The proposed design maintains the lowest MSE across
the considered range, demonstrating the benefit of jointly adapting
the PA positions and AirComp transceiver as the number of AirComp users increases.

\begin{figure}[]
    \centering
\includegraphics[width=0.9\linewidth, height=0.6\linewidth]{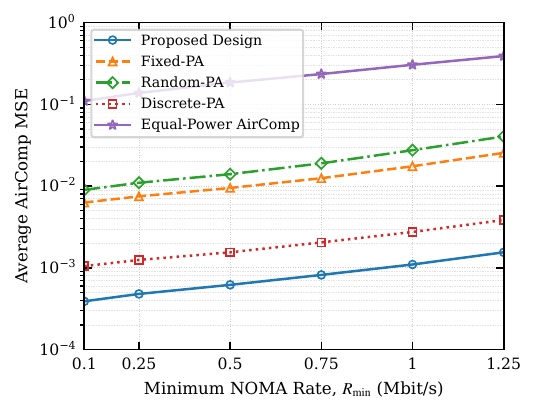}
    \caption{Average AirComp MSE versus the minimum NOMA rate requirement.}
    \label{fig:mse_vs_rate}
\end{figure}

Fig.~\ref{fig:mse_vs_rate} illustrates the impact of the minimum NOMA rate requirement
on the AirComp MSE. As $R^{\min}$ increases, the MSE increases for
all schemes since the more stringent NOMA QoS constraints reduce
the admissible AirComp interference level. The proposed design
consistently achieves the lowest MSE, demonstrating its ability to
better balance NOMA QoS requirements and AirComp accuracy through
joint PA-position and transceiver optimization. The performance
gaps over the benchmark schemes further highlight the benefits of
adaptive PA positioning and user-specific AirComp power control
under increasingly stringent communication requirements.

\begin{figure}[]
    \centering
\includegraphics[width=0.9\linewidth, height=0.6\linewidth]{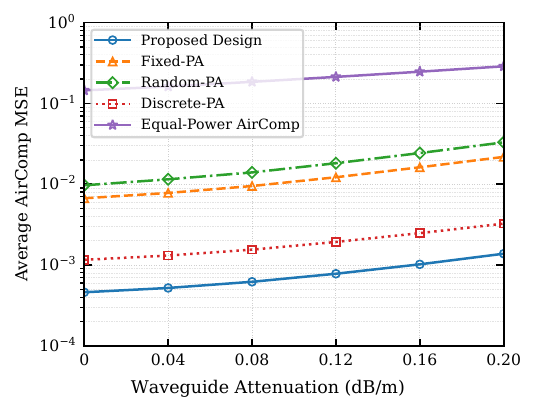}
    \caption{Average AirComp MSE versus the waveguide attenuation.}
    \label{fig:mse_vs_attenuation}
\end{figure}

Fig.~\ref{fig:mse_vs_attenuation} illustrates the impact of waveguide attenuation on the
AirComp MSE. As $\alpha_g$ increases, the MSE generally increases
due to the larger attenuation along the waveguide, which weakens
the effective channels and limits the flexibility for signal
alignment. The proposed design consistently achieves the lowest
MSE by adapting the PA positions and AirComp transceiver to the
attenuation-aware channel conditions. The performance gaps over
the benchmark schemes further demonstrate the importance of joint
PA-position and transceiver optimization in mitigating the impact
of waveguide attenuation.

\section{Conclusion}

This paper investigated a PAS-assisted integrated AirComp--NOMA
uplink with shared communication and computation resources. We
jointly optimized the AirComp transceiver, PA positions, and NOMA
SIC order to minimize the computation MSE under NOMA QoS,
transmit-power, and PA deployment constraints. By exploiting the
structure of the NOMA QoS constraints, we characterized NOMA
feasibility and the optimal SIC order, and developed an efficient
alternating optimization algorithm that generates a monotonically
nonincreasing, convergent MSE sequence. Numerical results
demonstrated that the proposed joint design substantially improves
computation accuracy over the considered benchmarks and effectively
balances AirComp performance with NOMA QoS requirements. These
results highlight the potential of PAS-enabled spatial
reconfigurability for efficient communication--computation
coexistence.

\appendices

\section{Proof of Proposition 1}
\label{app:proof_prop1}

Using \eqref{eq:noma_sinr} and \eqref{prob:P0_rate}, the QoS
constraint of user $\pi(m)$ can be equivalently written as
\begin{equation}
q_{\pi(m)}
\geq
\frac{\gamma_{\pi(m)}}{|g_{\pi(m)}(\mathbf v)|^2}
\left(
\sum_{\ell=m+1}^{K_{\rm N}}
|g_{\pi(\ell)}(\mathbf v)|^2 q_{\pi(\ell)}
+T
\right).
\label{eq:app_noma_power_recursion}
\end{equation}
For the last decoded user, \eqref{eq:app_noma_power_recursion} reduces
to
\begin{equation}
    q_{\pi(K_{\rm N})}
\geq
\frac{\gamma_{\pi(K_{\rm N})}T}
{|g_{\pi(K_{\rm N})}(\mathbf v)|^2},
\end{equation}
and hence its minimum feasible power is obtained by equality.
Since the right-hand side of
\eqref{eq:app_noma_power_recursion} is nondecreasing in the powers of
the subsequently decoded users, successively activating the QoS
constraints with equality in the reverse SIC order yields the
componentwise minimum feasible NOMA power vector. Applying backward
induction gives
\begin{equation}
    q_{\pi(m)}^{\min}
=
\frac{\gamma_{\pi(m)}T}
{|g_{\pi(m)}(\mathbf v)|^2}
\prod_{\ell=m+1}^{K_{\rm N}}
\left(1+\gamma_{\pi(\ell)}\right),
\end{equation}
which establishes \eqref{eq:min_noma_power}.

Therefore, every feasible NOMA power allocation satisfies
$q_{\pi(m)} \ge q_{\pi(m)}^{\min}$ for all $m$. Moreover, the
power vector defined by (15) satisfies all NOMA QoS constraints
with equality. Hence, a power allocation satisfying the individual
power budgets exists if and only if
$q_{\pi(m)}^{\min} \le P_{\pi(m)}$ for every $m$.
Substituting (15) into these inequalities yields (14), which
completes the proof.

\section{Proof of Proposition 2}
\label{app:proof_prop2}

Consider two adjacent NOMA users $i$ and $j$ in an arbitrary SIC
decoding order, and let
$C\triangleq\prod_{\ell\in\mathcal S}(1+\gamma_\ell)$, where
$\mathcal S$ denotes the set of users decoded after both $i$ and $j$.
Also define
$A_u\triangleq |g_u(\mathbf v)|^2P_u/\gamma_u$.
If $i$ is decoded immediately before $j$, the minimum of their two
terms in \eqref{eq:noma_tolerance} is
$C^{-1}\min\{A_i/(1+\gamma_j),A_j\}$.
If their positions are interchanged, the corresponding minimum is
$C^{-1}\min\{A_j/(1+\gamma_i),A_i\}$.

Suppose that $\rho_i(\mathbf v)\geq\rho_j(\mathbf v)$. From the
definition of the ordering metric, this condition is equivalent to
$A_i/(1+\gamma_j)\geq A_j/(1+\gamma_i)$.
Since $\gamma_i,\gamma_j>0$, we further have
$A_j>A_j/(1+\gamma_i)$ and
$A_i>A_i/(1+\gamma_j)\geq A_j/(1+\gamma_i)$.
It therefore follows that
\begin{equation}
\min\left\{
\frac{A_i}{1+\gamma_j},\,A_j
\right\}
\geq
\min\left\{
\frac{A_j}{1+\gamma_i},\,A_i
\right\}.
\label{eq:app_pairwise_order}
\end{equation}
Hence, whenever $\rho_i(\mathbf v)\geq\rho_j(\mathbf v)$, placing
$i$ immediately before $j$ does not decrease
$\Theta_{\boldsymbol{\pi}}(\mathbf v)$.

Moreover, interchanging two adjacent users leaves the tolerance terms
of all users decoded after them unchanged. The terms associated with
all users decoded before them are also unchanged, since they contain
the same factor $(1+\gamma_i)(1+\gamma_j)$ regardless of the relative
order of $i$ and $j$. Consequently, any adjacent inversion with
respect to the nonincreasing order of $\rho_j(\mathbf v)$ can be
removed without decreasing $\Theta_{\boldsymbol{\pi}}(\mathbf v)$.
By successively removing all such inversions, any SIC decoding order
can be transformed into a permutation satisfying
\eqref{eq:optimal_sic_order} without decreasing the NOMA interference
tolerance. Therefore,
$
\Theta_{\boldsymbol{\pi}^{\star}(\mathbf v)}(\mathbf v)
=
\max_{\boldsymbol{\pi}\in\Pi(\mathcal K_{\rm N})}
\Theta_{\boldsymbol{\pi}}(\mathbf v),
$
which establishes \eqref{eq:max_noma_tolerance} and completes the
proof.

\section{PA-Position Update}
\label{app:pa_majorization}

Define $\kappa\triangleq2\pi/\lambda$ and
$\beta_g\triangleq2\pi/\lambda_g$. The contribution of the $n$th PA
to the equivalent channel of user $u$ is
\begin{equation}
\psi_{u,n}(v_n)
\triangleq
\frac{\lambda}{4\pi d_{u,n}(v_n)}
e^{-\alpha_g v_n}
e^{-j[\kappa d_{u,n}(v_n)+\beta_g v_n]},
\end{equation}
such that $g_u(\mathbf v)=\sum_{n=1}^{N}\psi_{u,n}(v_n)$. Since
$\partial d_{u,n}/\partial v_n=(v_n-x_u)/d_{u,n}$, direct
differentiation yields
\begin{equation}
\frac{\partial g_u(\mathbf v)}{\partial v_n}
=
\psi_{u,n}(v_n)
\left[
-\frac{v_n-x_u}{d_{u,n}^2}
-\alpha_g
-j\kappa\frac{v_n-x_u}{d_{u,n}}
-j\beta_g
\right].
\label{eq:channel_position_derivative}
\end{equation}
Consequently,
$\partial |g_u(\mathbf v)|^2/\partial v_n
=2\operatorname{Re}\{g_u^*(\mathbf v)
\partial g_u(\mathbf v)/\partial v_n\}$.

Let $e_k(\mathbf v)\triangleq w^*p_k g_k(\mathbf v)-1$. Using
\eqref{eq:channel_position_derivative}, the gradient components
required in the quadratic models are
\begin{subequations}
\label{eq:pa_gradients}
\begin{align}
\frac{\partial f(\mathbf v)}{\partial v_n}
={}&
2\operatorname{Re}\left\{
\sum_{k\in\mathcal K_{\rm A}}
e_k^*(\mathbf v)w^*p_k
\frac{\partial g_k(\mathbf v)}{\partial v_n}
\right\},
\label{eq:pa_objective_gradient}
\\
\frac{\partial c_m^{(t)}(\mathbf v)}{\partial v_n}
={}&
2\operatorname{Re}\Bigg\{
\sum_{k\in\mathcal K_{\rm A}}
|p_k|^2g_k^*(\mathbf v)
\frac{\partial g_k(\mathbf v)}{\partial v_n}
\nonumber\\
&\hspace{7mm}
-a_m^{(t)}
g_{\pi^{(t)}(m)}^*(\mathbf v)
\frac{\partial g_{\pi^{(t)}(m)}(\mathbf v)}
{\partial v_n}
\Bigg\}.
\label{eq:pa_constraint_gradient}
\end{align}
\end{subequations}
Since $H>0$, all user--PA distances are bounded away from zero.
Thus, $f(\mathbf{v})$ and $c_m^{(t)}(\mathbf{v})$ are smooth
with bounded Hessians over the compact feasible PA region, and
therefore have Lipschitz-continuous gradients. Hence, by the descent lemma, finite positive curvature
parameters exist for which the quadratic models
\eqref{eq:pa_objective_majorizer} and
\eqref{eq:pa_constraint_majorizer} provide valid upper bounds.

The curvature parameters are determined by backtracking. For given
$L_f^{(t)}$ and $\{L_m^{(t)}\}$, Problem~\eqref{prob:pa_sca} is
solved to obtain a candidate $\widetilde{\mathbf v}$, which is
accepted if
$
f(\widetilde{\mathbf v})
\leq\widehat f_t(\widetilde{\mathbf v})$,
 $c_m^{(t)}(\widetilde{\mathbf v})
\leq\widehat c_{m,t}(\widetilde{\mathbf v}),
 m=1,\ldots,K_{\rm N}.$
Otherwise, the corresponding curvature parameters are increased and
the QCQP is solved again. The existence of finite Lipschitz constants
ensures termination of the backtracking procedure.

Finally, the surrogate functions are tight at $\mathbf v^{(t)}$.
Thus, a feasible $\mathbf v^{(t)}$ is also feasible for
\eqref{prob:pa_sca}. For every accepted update,
$
c_m^{(t)}(\mathbf v^{(t+1)})
\leq
\widehat c_{m,t}(\mathbf v^{(t+1)})
\leq0,
 m=1,\ldots,K_{\rm N},
$
which preserves feasibility for the current SIC order. Moreover,
$
f(\mathbf v^{(t+1)})
\leq
\widehat f_t(\mathbf v^{(t+1)})
\leq
\widehat f_t(\mathbf v^{(t)})
=
f(\mathbf v^{(t)}),
$
where the second inequality follows from the optimality of
$\mathbf v^{(t+1)}$ in \eqref{prob:pa_sca}. Hence, the PA-position
update preserves NOMA feasibility and does not increase the AirComp
objective.

\bibliographystyle{IEEEtran}
\bibliography{refpinching}

\vfill
	
\end{document}